\documentclass[epj]{svjour}
\usepackage{graphics}
\usepackage{axodraw}
\input{psfig.sty}
\usepackage{graphicx}% Include figure files
\usepackage{dcolumn}% Align table columns on decimal point

\newcommand{\ar}{\arrowvert}

\newcommand{\be}{\begin{equation}}
\newcommand{\ee}{\end{equation}}
\newcommand{\ba}{\begin{eqnarray}}
\newcommand{\ea}{\end{eqnarray}}

\begin{document}
\title{Heat conductivity of a pion gas}
\author{ Antonio Dobado Gonz\'alez, Felipe J. Llanes-Estrada and Juan
M. Torres Rinc\'on 
}                     % Do not remove
\institute{ Departamento de F\'{\i}sica Te\'orica I,  Universidad
Complutense, 28040 Madrid, Spain }
\date{Received: date / Revised version: date}

\abstract{ 
We evaluate the heat conductivity of a dilute pion gas employing the
Uehling-Uehlenbeck equation and experimental phase-shifts parameterized
by means of the $SU(2)$ Inverse Amplitude Method. Our results are 
consistent with previous evaluations. For comparison we also give 
results for an (unphysical) hard sphere gas.
 % end of PACS codes
} %end of abstract
\maketitle

%%%%%%%%%%%%%%%%%%%%%%%%%%%%%%%%%%%%%%%%%%%%%%%%%%%%%%%%%%%%%%%%%%%
\section{Transport coefficients}
%%%%%%%%%%%%%%%%%%%%%%%%%%%%%%%%%%%%%%%%%%%%%%%%%%%%%%%%%%%%%%%%%%%

Transport coefficients in heavy-ion collisions remain of interest as 
prospects for direct experimental measurement at RHIC improve. In 
particular, elliptic flow has been proposed \cite{Teaney:2003kp} as a 
tell-tale for viscous effects.
There have also recently been a number of papers emphasizing viscosity 
(for example \cite{Liu:2006cs}, \cite{Nakano:2006wy}) following recent 
insight from conformal 
field theory \cite{Policastro:2001yc}. Follow-up studies are 
under preparation by several groups from the hadron and from the 
quark-gluon phases in RHIC-theory to see the effect of the phase 
transition on the viscosity.

Although we have presented a comprehensive study of the shear viscosity 
of a hadronic gas at low temperature \cite{Dobado:2003wr},  in this 
brief report we focuse on another transport coefficient, the thermal 
conductivity, associated to energy conservation and flow in the hot gas. 

We employ the same notation and conventions as in our earlier 
publication \cite{Dobado:2003wr}. In particular we employ the $SU(2)$ 
Inverse Amplitude Method parametrization of the pion-pion scattering 
experimental phase shifts (as well as alternative parametrizations to 
check the sensitivity of the calculation presented). Note we also employ 
the Landau convention for the flows as opposed to the Eckart convention 
that has also been widely used \cite{Gavin:1985ph}.

To make this paper minimally self-contained, we collect here a few of 
the key results and assumptions.

We are working in a pion gas at temperatures well below any phase 
transition, in the dilute approximation near equilibrium. The 
equilibrium equation of state yields an enthalpy per unit 
volume $h=P+\rho$ as
\begin{equation}
h=\frac{g_{\pi}m_{\pi}^4}{4\pi^2} \int_0^\infty dx \frac{\sqrt{x} (1+4x/3)}{
\sqrt{x+1}\left( z^{-1} e^{y(\sqrt{1+x}-1)} -1 \right)}
\end{equation}
and the particle (pion) density is given by
\begin{equation}
n\left( y=\frac{m_{\pi}}{T},z \right) = \frac{g_{\pi}m_{\pi}^3}{4\pi^2}
\int_0^{\infty} dx \frac{x^{1/2}}{z^{-1} e^{y(\sqrt{1+x}-1)}
-1} 
\end{equation}
($z=e^{(\mu-m)/T}$ being the fugacity).

We take a constant perturbation around equilibrium proportional to the 
relativistic temperature-pressure gradient 
\be
\delta f ({\bf p})= \frac{f_0}{T} \frac{{\bf p}}{\ar{\bf p}\ar} 
g(\ar{\bf p}\ar)
\left[
{\bf \nabla} T - \frac{T}{h} {\bf \nabla}P\right]
\ee
and 
change the variable to $x=(p/m_\pi)^2$.
We perform a polynomial expansion of $g(x)$, and display results for the 
first order only, $g(x)=A_0$,  with $A_0$ a constant. Convergence is 
known to be fast.  
Since pions are not exactly massless, no singularities appear that 
might require a more careful treatment \cite{Manuel:2004iv} and the 
only reason to improve on this variational calculation would be to 
improve its accuracy beyond the 10\% level. This program has been 
carried out at least for the non-relativistic pion gas viscosity in ref. 
\cite{Dobado:2001jf}. We have also made minimum sensitivity checks based 
on a simple numeric Gram-Schmidt polynomial construction beyond $A_0$ 
but will report them elsewhere. 

From the standard linearized (Boltzmann) Uehling-Uehlenbeck transport 
equation one can derive the following for this perturbation near equilibrium:
\ba \label{supercalculo}
A_0= \frac{I_1(y,z)}{I_2(y,z)} \\
I_1=2\pi m_{\pi}^3 y \int_0^\infty \frac{xdx}{\sqrt{1+x}}
f(E(x)) \left( \sqrt{1+x} - \frac{h}{m_{\pi}n} \right) \\ 
I_2= \frac{1}{4A^2} \int [ d\Phi ] \left[ e^{\beta ( 
\omega - 2 \mu)} ff'f_1f'_1  \right. \\ \nonumber \left.
\Delta_{i} \left( {\bf 
\hat{p}_{i}}\left[1-e^{-\beta(E_{i}-\mu)}\right]\right) \cdot
\Delta_{j} \left( {\bf 
\hat{p}_{j}}\left[1-e^{-\beta(E_{j}-\mu)}\right]\right) \right]
\ea
where $\omega$ denotes the total energy ($\omega=E+E_{1}=E'+E'_{1}$) ,
$A$ is the inverse normalization constant of the distribution functions
$$ A=\xi_{\pi}^{-1}= \frac{g_{\pi}}{(2 \pi) ^{3}}$$
($g_\pi=3$ for isospin degeneracy)
and the last term is a shorthand for the symmetrization
\ba 
\Delta_{i} \left( {\bf \hat{p}_{i}} \left[1-e^{-\beta(E_{i}-\mu)} 
\right] 
\right) =  \\ \nonumber
\left( {\bf \hat{p}'_{1}} \left[ 1-e^{-\beta(E'_{1}-\mu)} \right] 
+ {\bf \hat{p}'} \left[ 1-e^{-\beta(E'-\mu)} \right]-  
\right.\\ \left. \nonumber
{\bf \hat{p}_{1}} \left[ 1-e^{-\beta(E_{1}-\mu)} \right]- 
{\bf \hat{p}} \left[ 1-e^{-\beta(E-\mu)} \right] \right) \ .
\ea
Once $A_0$ has been thus computed, the heat conductivity follows as
\be
\kappa= -\frac{h m_{\pi}^3 2\pi A}{3nT} A_0 l_{1}(y,z)
\ee
with 
\be
l_{1}(y,z)= \int_0^\infty \frac{xdx }{\sqrt{1+x}
\left( z^{-1}e^{y(\sqrt{1+x}-1)}-1\right)}
\ee
To evaluate the integral in $I_2$ in Eq. (\ref{supercalculo}) we employ 
a Montecarlo computer program.
Without loss of generality we can choose the total momentum directed 
along the $OZ$ axis.
The independent variables can be taken as $P$ and $\omega$ 
(respectively the total momentum and energy in a binary collision), 
$p=\ar {\bf p}\ar$ and $p'=\ar {\bf p}'\ar$  (the incoming and outgoing 
pion momenta for one of the two pions, the other being fixed by momentum 
conservation). Finally the outgoing pion with momentum $p'$ does not 
need to be in the same plane as ${\bf P}$ and ${\bf p}$, and therefore 
we need an azimuthal angle $\phi'$ for this momentum. 
The cosines of the polar angles associated to ${\bf p}$ and $\bf{p'}$ 
are fixed by the energy conservation relation
$$\delta \left( \omega-E-E_{1} \right) = \frac{E_{1}}{p P} \,
\delta (x-x_{0})$$
and the associated integrals are immediately performed, with
\begin{eqnarray}
x_0= \frac{P^2 + \omega (2E-\omega)}{2pP} \\ 
x_0'=\frac{P^2 + \omega (2E'-\omega)}{2p'P} \ .
\end{eqnarray}
In addition, rigid rotations around ${\bf P}$ parametrized by $\phi$ are 
trivial and lead to a factor of $2\pi$, and global rotations of the 
system (the angles associated to ${\bf P}$) are also trivial and yield 
another $4\pi$.

Putting all together, the phase space integration \\
weighted with the 
square scattering amplitude can be expressed as
\ba
\int [d\Phi]=\int v_{rel} d\sigma d{\bf p} d{\bf p}'=
\\ \nonumber
\int dP dp dp' d\omega d\phi' \ar T \ar^2 \frac{4\pi 2\pi}{4\pi^2}
\frac{1}{4EE_1} \frac{p'}{E'_1} \frac{p'}{E'} \frac{E_1}{pP}
\frac{E_1'}{p'P} P^{2} p^{2} \ 
\ea

Our numerical results for the thermal conductivity are displayed in the 
figures. 
\begin{figure}
\includegraphics[width=3in, angle=-90]{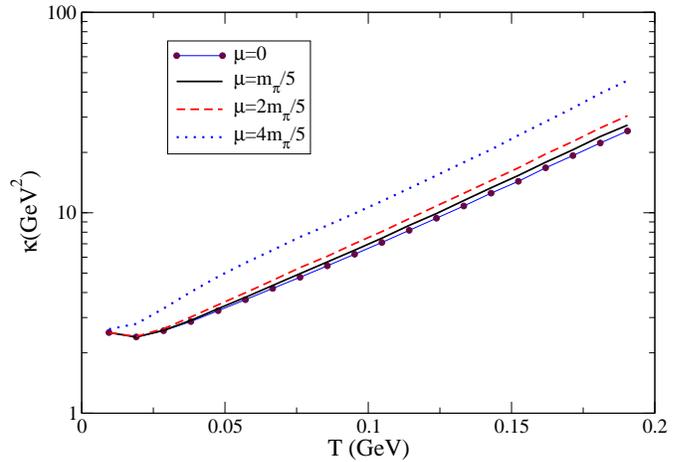}
\caption{We show the $\mu$ dependence of the conductivity 
for the hard-sphere gas appoximation}
\label{kamplitudfija}
\end{figure}

In Fig.~\ref{kamplitudfija} we plot the heat conductivity as a function 
of temperature, at several $\mu$'s in the hard-sphere scattering case, 
that is, when we use a constant scattering amplitude based on 
Weinberg's low energy theorem:
\be
\left|T\right|^2=\frac{23}{3} \frac{m_{\pi}^{4}}{f_{\pi}^{4}}
\ee
For no $\mu$ does  the heat conductivity diverge at low 
temperature to the reach of our Montecarlo.
In the high temperature limit we expect on 
dimensional grounds, and numerically find, $\kappa = A \cdot T^2$, 
where the numerical constant is close to $A=685$.

\begin{figure}
\includegraphics[width=3in,angle=-90]{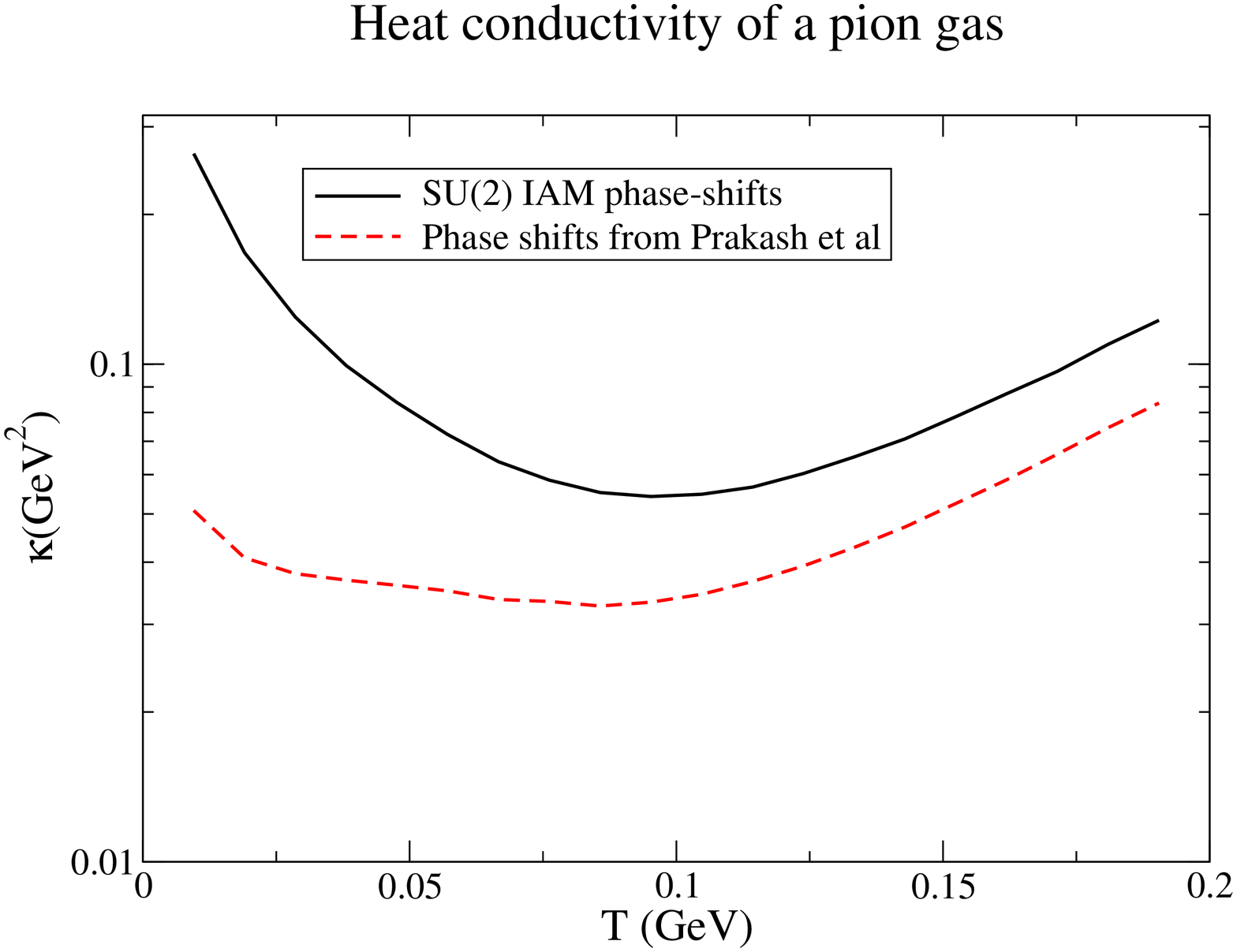}
\caption{Thermal conductivity $\kappa$ from SU(2) 
chiral perturbation theory unitarized by means of the Inverse Amplitude 
Method. The pion chemical potential is taken to be $\mu_\pi=0$.
\label{so2}}
\end{figure}
In Fig.~\ref{so2}, employing IAM phase shifts, that unitarize  a higher 
order $O(p^4)$ in chiral perturbation theory 
\cite{Dobado:1989qm}, and 
therefore introduce a dependence with the energy for the pion-pion 
scattering amplitude,  we now find that there is a minimum 
value near $T_{c}=100$ MeV and $\kappa \to \infty $ when $T\to 0$.
The high $T$ scaling is close to the dimensional analysis even at the 
moderately large plotted temperatures,
 $$\kappa \propto 3.4 \, T^2\ .$$
In the figure we also show our own Montecarlo-based calculation of the 
heat conductivity 
employing the simple resonance
saturation parametrization for the isoscalar  and isovector phase
shifts by the Brookhaven group 
\cite{Prakash:1993bt}, for comparison 
and to give an idea of the sensitivity to the parametrization.

\begin{figure}
\includegraphics[width=3in,angle=-90]{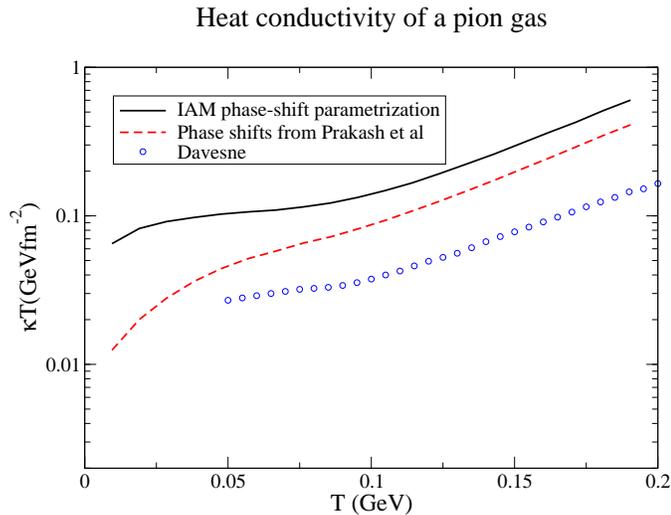}
\caption{
Thermal conductivity times temperature $T\kappa$ from SU(2)
chiral perturbation theory unitarized by means of the Inverse Amplitude
Method. The pion chemical potential is taken to be $\mu_\pi=0$. We 
also plot the computation with the phase-shifts of \cite{Prakash:1993bt}
and compare them to D. Davesne's evaluation.
\label{comparacion}}
\end{figure}

Next in Fig.~\ref{comparacion} we show the comparison with the existing
computation of D. Davesne \cite{Davesne:1995ms} of both calculations 

Note also a recent calculation by \cite{Muroya:2004pu} that 
employs chiral perturbation theory alone (without unitarization). 
This approach features an ever-increasing 
cross section, unphysical behavior that artificially shortens the mean 
free path and therefore lowers the transport coefficients. Therefore the 
validity of the results is limited to low temperatures. 
However a direct comparison with this approach is difficult since the 
authors directly include the effect of baryon resonances.

In conclusion, from published calculations and our own contribution we 
see that we have a fair theoretical idea on how the thermal 
conductivity of a pure pion gas 
should behave with temperature. We have further studied its behavior 
with chemical potential that is important because chemical freeze-out is 
expected to occur before thermal freeze out in Heavy-Ion Collisions, the 
reason being that the low energy hadron interactions are elastic 
scattering, largely mediated by resonances (automatically incorporated 
in our Inverse Amplitude Method Phase shifts).

\vspace{2cm}
{\it{
This work has been supported by research grants FPA 2004-02602, 
2005-02327, PR27/05-13955-BSCH and has been presented at the IVth
International Conference on Quarks and Nuclear Physics, Madrid, Spain, 
June 5th-10th 2006.
We thank A. G\'omez Nicola and D. Fern\'andez-Fraile for informing us 
that their own calculation along the lines of \cite{Fernandez-Fraile:2006sk}
in chiral perturbation theory also yields analogous results.
}}


\begin{thebibliography}{99}

\bibitem{Teaney:2003kp}
  D.~Teaney,
  %``Effect of shear viscosity on spectra, elliptic flow, and Hanbury
  %Brown-Twiss radii,''
  Phys.\ Rev.\  C {\bf 68} (2003) 034913
  [arXiv:nucl-th/0301099].
  %%CITATION = PHRVA,C68,034913;%%

\bibitem{Liu:2006cs}
  H.~Liu, D.~Hou and J.~Li,
  %``Shear viscosity and viscous entropy production in hot QGP at finite
  %density,''
  arXiv:hep-ph/0609034.
  %%CITATION = HEP-PH/0609034;%%

\bibitem{Nakano:2006wy}
  E.~Nakano,
  %``Shear viscosity of pion gas,''
  arXiv:hep-ph/0612255.
  %%CITATION = HEP-PH/0612255;%%

\bibitem{Policastro:2001yc}
  G.~Policastro, D.~T.~Son and A.~O.~Starinets,
  Phys.\ Rev.\ Lett.\  {\bf 87} (2001) 081601
  [arXiv:hep-th/0104066].
  %%CITATION = PRLTA,87,081601;%%

\bibitem{Dobado:2003wr}
  A.~Dobado and F.~J.~Llanes-Estrada,
  %``The viscosity of meson matter,''
  Phys.\ Rev.\  D {\bf 69} (2004) 116004
  [arXiv:hep-ph/0309324].
  %%CITATION = PHRVA,D69,116004;%%

\bibitem{Gavin:1985ph}
  S.~Gavin,
  %``Transport Coefficients In Ultrarelativistic Heavy Ion Collisions,''
  Nucl.\ Phys.\  A {\bf 435} (1985) 826; \\
  %%CITATION = NUPHA,A435,826;%%
  W. A. van Leeuwen and S. R. de Groot, Physica {\bf 51} (1971) 1. 

\bibitem{Manuel:2004iv}
  C.~Manuel, A.~Dobado and F.~J.~Llanes-Estrada,
  %``Shear viscosity in a CFL quark star,''
  JHEP {\bf 0509} (2005) 076
  [arXiv:hep-ph/0406058].
  %%CITATION = JHEPA,0509,076;%%

\bibitem{Dobado:2001jf}
  A.~Dobado and S.~N.~Santalla,
  %``Pion gas viscosity at low temperature and density,''
  Phys.\ Rev.\  D {\bf 65} (2002) 096011
  [arXiv:hep-ph/0112299].
  %%CITATION = PHRVA,D65,096011;%%

\bibitem{Dobado:1989qm}
  A.~Dobado, M.~J.~Herrero and T.~N.~Truong,
  Phys.\ Lett.\  B {\bf 235} (1990) 134.
  %%CITATION = PHLTA,B235,134;%%

\bibitem{Prakash:1993bt}
  M.~Prakash, M.~Prakash, R.~Venugopalan and G.~Welke,
  %``Nonequilibrium properties of hadronic mixtures,''
  Phys.\ Rept.\  {\bf 227} (1993) 321.
  %%CITATION = PRPLC,227,321;%%

\bibitem{Davesne:1995ms}
  D.~Davesne,
  %``Transport coefficients of a hot pion gas,''
  Phys.\ Rev.\  C {\bf 53} (1996) 3069.
  %%CITATION = PHRVA,C53,3069;%%

\bibitem{Muroya:2004pu}
  S.~Muroya and N.~Sasaki,
  Prog.\ Theor.\ Phys.\  {\bf 113} (2005) 457
  [arXiv:nucl-th/0408055].
  %%CITATION = PTPKA,113,457;%%

\bibitem{Fernandez-Fraile:2006sk}
  D.~Fernandez-Fraile and A.~G\'omez-Nicola,
  %``Transport coefficients in chiral perturbation theory,''
  arXiv:hep-ph/0610197.
  %%CITATION = HEP-PH/0610197;%%

\end{thebibliography}
\end{document}